\documentclass[twocolumn,prl,showpacs]{revtex4}
\usepackage{graphicx}
\usepackage{psfrag}
\usepackage{color}
\usepackage{subfigure}
\usepackage{amsmath}

\def\avg#1{\left\langle#1\right\rangle}

\def\be{\begin{equation}}       \def\ee{\end{equation}}
\def\bea{\begin{eqnarray}}      \def\eea{\end{eqnarray}}
\def\ba{\begin{array} }
\def\ea{\end{array} }
\def\bnum{\begin{enumerate} }
\def\enum{\end{enumerate}}

\def\=>{\Rightarrow}
\def\>{\rightarrow}

\def\eye2{Fathbb{I}}

\renewcommand{\>}{\rangle}

\begin{document}
\title{Character of frustration on magnetic correlation in doped Hubbard model}
\author{Peng Wang$^{1}$, Xinran Ma$^{1}$, Jingyao Wang$^{1}$, Yamei Zeng$^{1}$, Ying Liang$^{1}$\footnote{liang@bnu.edu.cn},
and Tianxing Ma$^{1,2}$}
\affiliation{$^{1}$Department of Physics, Beijing Normal University, Beijing 100875, China%
\\
$^{2}$ Department of Physics, University of California, San Diego, California 92093, USA}

\date{\today}

\begin{abstract}
 The magnetic correlation in the Hubbard model on a two-dimensional anisotropic triangular lattice is studied by using the determinant quantum Monte Carlo method.
  Around half filling, it is found that the increasing frustration $t'/t$ could change the wave vector of maximum spin correlation along ($\pi,\pi$)$\rightarrow$($\pi,\frac{5\pi}{6}$)$\rightarrow$($\frac{5\pi}{6},\frac{5\pi}{6}$)$\rightarrow$ ($\frac{2\pi}{3},\frac{2\pi}{3}$), indicating the frustration's remarkable effect on the magnetism. In the studied filling region $\avg{n}=1.0-1.3$, the doping behaves like some kinds of {\it{frustration}}, which destroys the $(\pi,\pi)$ AFM correlation quickly and push the magnetic order to a wide range of the $(\frac{2\pi}{3},\frac{2\pi}{3})$ $120^{\circ}$ order when the  $t'/t$  is large enough.
  Our non-perturbative calculations reveal a rich magnetic phase diagram over both the frustration and electron doping.
\end{abstract}


\maketitle

\section{Introduction}

Geometrically frustrated systems are among the central issues in condensed-matter
physics for many years\cite{Lacroix2011}. Geometrical frustration is caused by a lattice
structure where intersite interactions are conflicting. This conflict is reflected by the
impossibility of a simultaneous pairwise minimization of the interaction energy on a given
site and its neighbors. In geometrically frustrated systems, electron correlations may lead to exotic quantum
states and properties. Understanding such novel quantum phenomena will give new
insights in physics beyond particular fields. Recently, several new materials with frustrations have been discovered and novel
phenomena in these systems have been reported. Among them, experimental and theoretical studies have demonstrated that the anisotropic triangular lattice  as realized in the $\kappa$-(BEDT-TTF)$_{2}$X family of organic charge transfer salts yields a complex phase diagram with magnetic\cite{Bulut2005,Li2014,Yamada2014}, superconducting\cite{David2005}, Mott insulating\cite{Capone2001,Kyung2007,Phase2015}, and spin
liquid phases\cite{Elsinger2000,Shimizu2003,YKurosaki2005,Kagawa2005,Sahebsara2008,Yang2010}.

Finding
the underlying mechanism is a great challenge to physicists and
there are intense efforts to get a grip on the problem  including studies with path integral renormalization group\cite{Morita2002,Mizusaki2006,Yoshioka2009},
exact diagonalization\cite{Clay2008,Koretsune2007}, variational Monte Carlo
calculations\cite{Watanabe2006,Watanabe2008}, cluster dynamical mean field
theory\cite{Kyung2006,Ohashi2008}, and dual fermions \cite{Lee2008}.
Referring to the magnetism, $\kappa$-(ET)$_{2}$Cu[N(CN)$_{2}$]Cl exhibits an antiferromagnetic (AFM)
long range order in the insulating state\cite{Yasin2011,Lunkenheimer2012}, where $\kappa$-(ET)$_{2}$Cu(NCS)$_{2}$ has no magnetic order.  The situation of the
pairing symmetry in $\kappa$-(ET)$_{2}$X is also complicated owing to the existence of frustration.  Some studies on the anisotropic
triangular lattice and NMR experiments\cite{Mayaffre1995} suggested
that most of the $\kappa$--type salts have $d_{x^{2}-y^{2}}$-wave
pairing, while some other experiments for
$\kappa$-(ET)$_{2}$Cu(NCS)$_{2}$ have implied that the directions of
nodes are similar to that of $d_{xy}$-wave\cite{Izawa2001,Arai2001,Taylor2007}.

\begin{figure}[tbp]
\includegraphics[width=240pt]{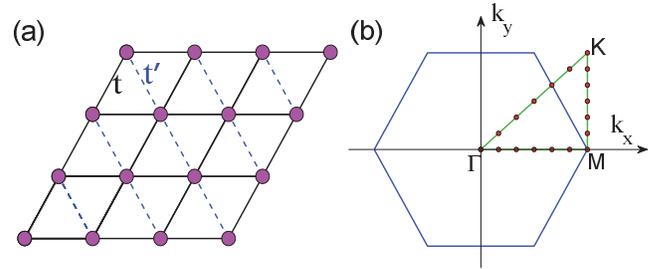}
\caption{(Color online) (a) The anisotropic triangular lattice and (b) the first Brillouin zone (the red points along the green line represent the high symmetry points including $\Gamma(0,0)$, $M(\pi,0)$ and $K(\pi,\pi)$ points). Solid and dashed lines denote the hopping amplitudes $t$ and the $t'$, respectively.}
\label{Fig:Sketch}
\end{figure}

The complexity of these family members calls for further studies on
the anisotropic triangular lattice so as to clarify some important
issues. The anisotropic triangular lattice is an ideal platform to study the effect of frustration. As demonstrated in Fig. \ref{Fig:Sketch} (a), $t$ denotes the nearest hoping and $t'$ indicates the next nearest hoping term.  By varying the ratio $t'/t$ from 0 to 1, the systems go from
non-frustrated ( $t'$ =0) to fully-frustrated ones ( $t'$ =1). In fact, the change of $t'/t$  is not only a way to control frustration, but also describes different real materials.

%

The geometrical frustration measured by $t'/t$ is expected to play a crucial role in inducing various magnetic order and superconducting states in geometrically frustrated systems. Corresponding to the non-frustrated cases ( $t'/t$=0 ), the ground state of the square lattice at half filling is a Mott insulator with AFM order.
Introducing frustration, which brings the competition between interactions and the geometry, is believed to suppress AFM ground state and replace it with other kinds of magnetic order\cite{McKenzie1998,Miyagawa1995,Kanoda2011,Struck2011,Gan2005,Gan2006,Kashima2001,Morita2002}.
Indeed, previous studies of Heisenberg model have suggested AFM N\'{e}el order for small frustration, incommensurate $(q,q)$ long-range AFM order for larger frustration,  and the  $120^{\circ}$ order realized in the isotropic limit\cite{Weihong1999,Chung2001,Yunoki2006,Powell2007,Ohashi2008,Scriven2012}.
%
However, most of these studies are limited on the half-filled case or finite frustrated case\cite{Kyung2006,Sahebsara2006,Watanabe2006,Clay2012}, and a full diagram on the magnetic order depends on the frustration and doping is still lacking, especially within some kinds of non-biased numerical methods. In this paper, we studied the magnetic correlation in the Hubbard model on a two-dimensional anisotropic triangular lattice by using the determinant quantum Monte Carlo method (DQMC). Our non-perturbative calculations reveal a rich magnetic phase diagram over both the frustration and doping.

\section{Model and Methods}

We start from the Hubbard model on the anisotropic
triangular lattice, which is defined as
\begin{equation}
\begin{aligned}
  \mathcal{H}\;=\;&t\sum_{\mathbf{\langle i,j\rangle }\sigma}(c_{\mathbf{i}\sigma }^{\dag }c_{\mathbf{j}\sigma }+h.c.)+t^{\prime}\sum_{\mathbf{\langle k,l\rangle }\sigma}(c_{\mathbf{k}\sigma }^{\dag}c_{\mathbf{l}\sigma}+h.c.)\\
  &+U\sum_{\mathbf{i}}n_{\mathbf{i}\uparrow}n_{\mathbf{i}\downarrow}
  -\mu\sum_{\mathbf{i}\sigma }n_{\mathbf{i}\sigma}
\end{aligned}
\label{Eq1}
\end{equation}
In Eq. (\ref{Eq1}), $c_{\mathbf{i}\sigma}$ ($c_{\mathbf{i}\sigma}^{\dagger}$) annihilates (creates) electrons at the site $R_{\mathbf{i}}$ with spin $\sigma$ ($\sigma=\uparrow,\downarrow$) and $n_{\mathbf{i}\sigma}=c^{\dagger}_{\mathbf{i}\sigma}c_{\mathbf{i}\sigma}$. Here $U$ is the onsite repulsion, $t$ and $t'$ represents the hopping amplitudes between nearest neighbors and next-nearest neighbors on the two-dimensional square lattice, as that illustrated in Fig. \ref{Fig:Sketch}.

By controlling the scale of the frustration parameter $t'/t$ $(0\sim 1.0)$, the anisotropic triangular lattice becomes a unified description of lattices between the square lattice ($t'/t=0$) and the isotropic triangular lattice ($t'/t=1$). At $t'=0$, neighboring lattice points have opposite spins so the system could maintain stable. As $t'/t$ increase, the model will show the characteristics of anisotropic triangular lattices. Experimental results have found that $t'/t$ lies in the range $0.4\lesssim t'/t\lesssim 0.84$ for $\kappa$-(BEDT-TTF)$_{2}$X family\cite{Kandpal2009,Nakamura2009}.


To study the magnetic correlations, we define the spin structure factor in the z direction at zero frequency,
\begin{eqnarray}
S(\textbf{q}) = \frac{1}{N_{s}} \sum_{\mathbf{i},\mathbf{j}} e^{i\textbf{q}\cdot(\textbf{R}_{\mathbf{i}}-\textbf{R}_{\mathbf{j}})}\langle m^{z}(R_{i})\cdot m^{z}(R_{j})\rangle
\end{eqnarray}
where $m^{z}(R_{i})=c_{i\uparrow}^{\dagger}c_{i\uparrow}-c_{i\downarrow}^{\dagger}c_{i\downarrow}=n_{\uparrow}(R_{i})$, $m^{z}(R_{j})=c_{j\uparrow}^{\dagger}c_{j\uparrow}-c_{j\downarrow}^{\dagger}c_{j\downarrow}=n_{\uparrow}(R_{j})$, and $N_{s}$ represents the unit number of the lattice.

Our main method is the DQMC simulation based on Blankenbecler-Scalapino-Sugar (BSS) algorithm\cite{Blankenbecler1981}, which is a reliable tool for
investigating the nature of magnetic correlations in the presence of moderate Coulomb interactions\cite{Txma2010,Txma2011}.
In DQMC, the basic strategy is to express the partition function as a high-dimensional integral over a set of
random auxiliary fields. Then the integral is accomplished by Monte Carlo techniques. In present simulations,
8000 sweeps were used to equilibrate the system, and an additional more than 30000 sweeps were made, each of which generated a
measurement. These measurements were split into ten bins which provide the basis of coarse-grain averages, and errors were estimated based on standard
deviations from the average. For more technique details, we refer to reference \cite{Blankenbecler1981}, and a brief introduction to the DQMC method based
on the BSS algorithm has been given in references. \cite{Txma2011,Raimundo2003}.

\section{Results and discussion}

%

\begin{figure}
  \centering
  \includegraphics[width=240pt]{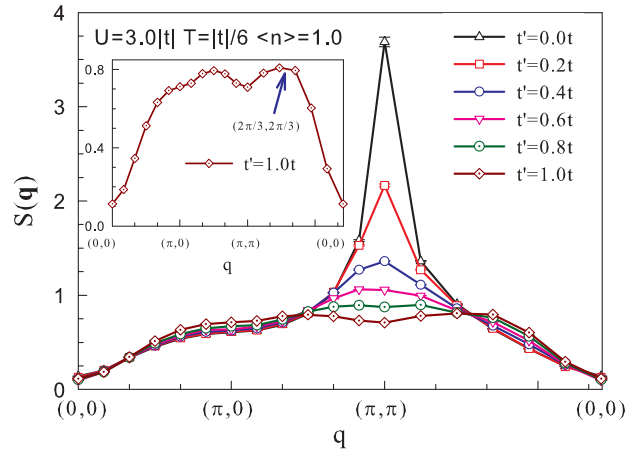}
  \caption{(Color online) Spin structure factor $S(\textbf{q})$ versus the momentum \textbf{q} for various frustration parameter $t'/t$ (ranged from $t'=0$ to $t'/t=1.0$) at $\frac{1}{2}$-filling on a $12^{2}$ lattice with $U=3.0|t|$, $T=|t|/6$. Here, \textbf{q} is scanned along the path $(0,0)\rightarrow(\pi,0)\rightarrow(\pi,\pi)\rightarrow(0,0)$ in trigonal first Brillouin zone.}
\label{Fig:Spin}
\end{figure}
Our DQMC simulations of the system were mainly performed at finite temperatures on a $12 \times 12$ lattices with periodic boundary conditions, and
our results are limited up to $U\leq 5.0|t|$ and $\beta=1/T \leq 6|t|$, due to the notorious sign problem in DQMC method. The band width of anisotropic triangle lattice is
in the range of $(8|t|,9|t|)$. The investigated interaction is around the half band width, for which the system is moderate correlated but not a strong correlated problem. Even so, the DQMC runs has to be stretched by a factor on the order of $\avg{sign}^{-2}$ for some $t'$ or electron fillings, in order
to obtain the same quality of data as for $\avg{sign}\simeq 1$.

In Fig. \ref{Fig:Spin}, the spin structure factor $S(\textbf{q})$ as a function of momentum $\textbf{q}$ with different $t'/t$ at half filling is present for $U=3.0|t|$ and temperature $T=|t|/6$. As the frustration, $t'/t$, is increased from 0 to 1.0, the results shown in Fig. \ref{Fig:Spin}, describe a whole diagram over the complete range from
the non-frustrated square lattice to the full-frustrated triangular lattice.  Our main aim here is to explore the role of magnetic order  $S(\textbf{q})$ depends on the frustration.
At $t'/t=0$ (dark line with triangle), a sharp peak of $S(\textbf{q})$ appears at the ($\pi,\pi$) point, indicating the intensive AFM fluctuation in the non-frustrated cases, and when $t'/t=1$  ( dark-red line with dotted diamond), the peak of $S(\textbf{q})$ moves to ($\frac{2\pi}{3},\frac{2\pi}{3}$), characteristic of the $120^{\circ}$ ordering, which has been proved in a series of previous references \cite{Clay2012,Powell2007,Watanabe2006,Yoshioka2009}.
In the range of  $0.6 < t'/t < 1.0$,  the peak of $S(\textbf{q})$  varies from the ($\pi,\pi$) point to ($\pi,\frac{5\pi}{6}$), and then ($\frac{5\pi}{6},\frac{5\pi}{6}$).
Thus, it is interesting to see that the magnetic order  has a rich phase diagram over the ratio of $t'/t$.  If there was a phase transition, it should be second order as the chemical potential around the phase transition is linear dependent on
 the change of the driving parameter $t'$, as well as the electron filling reported later.

\begin{figure}
  \centering
  \includegraphics[width=240pt]{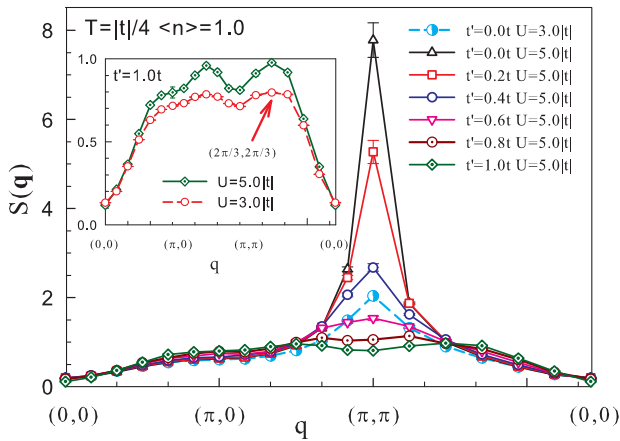}
  \caption{(Color online) Spin structure factor $S(\textbf{q})$ versus the momentum \textbf{q} for various frustration parameter $t'/t$ (ranged from $t'=0$ to $t'/t=1.0$) at $\frac{1}{2}$-filling on a $12^{2}$ lattice with $U=5.0|t|$, $T=|t|/4$.}
\label{Fig:Spinb}
\end{figure}

 We further our studies on different $U$ and temperature. In Fig. \ref{Fig:Spinb}, a similar phase diagram with a larger $U=5.0|t|$ and higher temperature $T=|t|/4$ is shown, which confirms that the magnetic order varies over the frustration. In Fig. \ref{Fig:Spinb}, we also compare the magnetic correlation  of $U=3.0|t|$ with $U=5.0|t|$ when $t'=0$ and $t'=t$ at the same temperature $T=|t|/4$. The magnetic correlation is enhanced by the repulsive interactions, especially at the ($\pi,\pi$) for $t'=0$  and ($\frac{2\pi}{3},\frac{2\pi}{3}$) for $t'=t$.

\begin{figure}
  \centering
  \includegraphics[width=240pt]{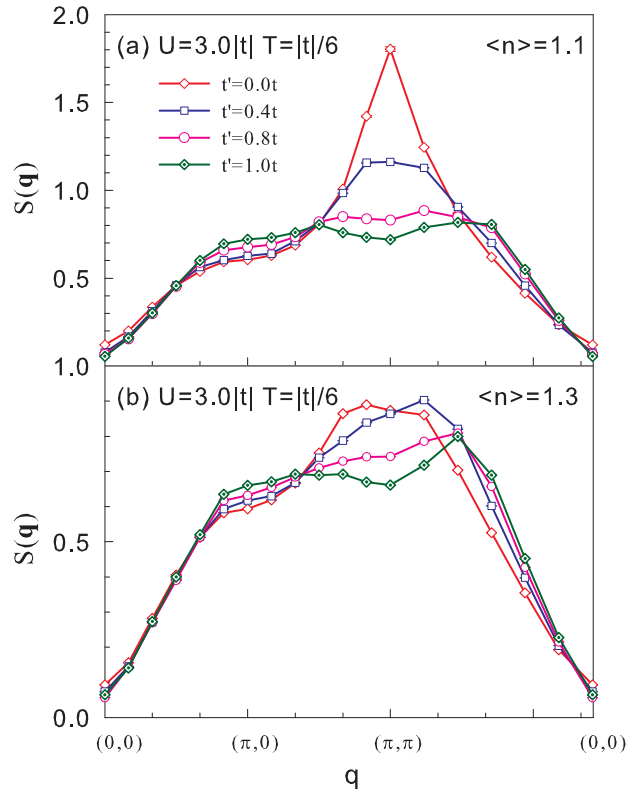}\\
  \caption{(Color online) Spin structure factor $S(\textbf{q})$ versus the momentum \textbf{q} for various frustration parameter $t'/t$ at the electron fillings (a) $\avg{n}=1.1$ and (b) $\avg{n}=1.3$, with $U=3.0|t|$ and $T=|t|/6$ on a $12^{2}$ lattice.}
\label{Fig:Spin13}
\end{figure}

In electronic correlated systems, doping is an important parameter, which may lead to new states. Fig. \ref{Fig:Spin13} shows the behaviors of $S(\textbf{q})$ at different $t'/t$ for (a) $\avg{n}=1.1$ and (b) $\avg{n}=1.3$ , respectively. Comparing Fig. \ref{Fig:Spin13} (a) with Fig. \ref{Fig:Spin}, it is seen that the peak of $S(\textbf{q})$ at $(\pi,\pi)$ decreases as $\avg{n}$ increases from 1.0 to 1.1, indicating the AFM fluctuations are suppressed by the increasing doping.
At $t'=0.4t$, the peak of $S(\textbf{q})$ varies from $(\pi,\pi)$ to $(\pi,\frac{5\pi}{6})$, and this crossover, requires a higher $t'/t$ ($0.6<(t'/t)<0.8$) for half filled case. The result shown in Fig. \ref{Fig:Spin13} (b) for $\avg{n}=1.3$ is similar as those for $\avg{n}=1.1$ and at half filling. When $t'/t$ increases from zero to $t'/t=0.4$, the peak of $S(\textbf{q})$ varies from $(\pi,\frac{5\pi}{6})$ to $(\frac{5\pi}{6},\frac{5\pi}{6})$, and finally locates at $(\frac{2\pi}{3},\frac{2\pi}{3})$ as $t'/t>0.8$.

It is known that the peak of $S(q)$ at $(\pi,\pi)$ indicates the AFM in square lattice ( non-frustrated system), and the peak of $S(q)$ at $\frac{2\pi}{3},\frac{2\pi}{3}$) corresponds to the 120-AFM in triangular lattice (full-frustrated system) at half fillings. For the magnetic wave vectors ($\pi,\frac{5\pi}{6}$) and ($\frac{5\pi}{6},\frac{5\pi}{6}$), we argue here that it is a crossover from the well format AFM to 120-AFM in the anisotropic triangle lattice. One interesting question to be asked here is whether there shall be long range magnetic order for phase with peak at $(\pi,\frac{5\pi}{6})$ or $(\frac{5\pi}{6},\frac{5\pi}{6})$.  To determine whether there is long range order, a careful scaling analysis with respect to different cluster size at zero temperature has to be done. However, the Hubbard model on an anisotropic triangular lattice is a strong frustrated system, and the notorious sign problem, prevent possible exact ground state projective quantum Monte carlo (PQMC) investigation. It is possible to try the constrained path Monte Carlo (CPMC) simulations, which uses the constrained path approximation, and it is free of any sign decay\cite{Zhangcpmc,Macpmc}. Further intensive numerical simulation is needed to clarify this point.

\begin{figure}
  \centering
  \includegraphics[width=240pt]{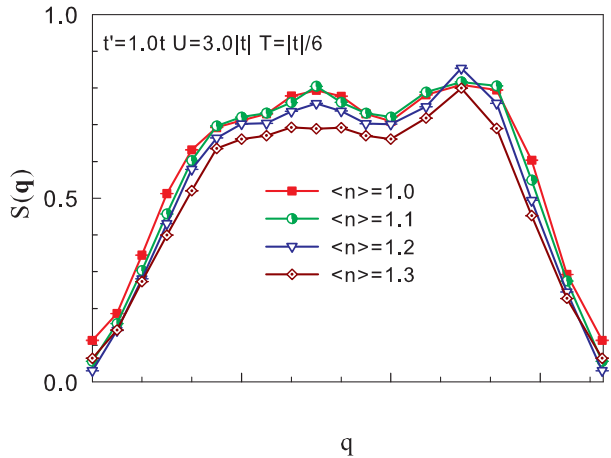}
  \caption{(Color online) Spin structure factor $S(\textbf{q})$ versus the momentum \textbf{q} at various electron fillings ( $\avg{n}$=1.0, 1.1, 1.2 and 1.3) with $t'= 1.0t$, $U=3.0|t|$ and $T=|t|/6$.}
\label{Fig:Spinn}
\end{figure}

 Fig. \ref{Fig:Spin13} reveals that the magnetic order may also have a rich phase diagram depends on the doping. To further our story step by step, $S(\textbf{q})$ at different fillings for $t'/t=1.0$ are shown in Fig. \ref{Fig:Spinn}. At half filling, fixed at the maximum frustration $t'/t=1$, a broad peak is around $(\frac{2\pi}{3},\frac{2\pi}{3})$. As the electron filling $\avg{n}$ increases from half filling to $\avg{n}=1.3$, $S(\textbf{q})$ in first Brillouin zone is reduced entirely except at $(\frac{2\pi}{3},\frac{2\pi}{3})$ where the changes of spin fluctuations are quite slight. Therefore, we could see that $S(\frac{2\pi}{3},\frac{2\pi}{3})$ remains the peak in the momentum space and the peak becomes much sharper as $\avg{n}$ increases. That seemingly shows the enhancement of $120^{\circ}$ ordering within the increase of certain electron doping.
 The enhancement of $120^{\circ}$ order at $\avg{n}=1.3$ may be
 understood as following.  Around $\avg{n}=4/3$, a 3-site triangle cluster is doped by one additional electron (i.e. 3 sites with 4 electrons). As this additional electron has equal possibility to stay at any of the three sites in one triangular cluster, to minimize the total energy of the spin interactions, the other three spins may tend to form a $120^{\circ}$  configuration. As in this case, all 3 sites are occupied by one electron and their spins point to each other with 120 angle, when the fourth electron is added, at each site its spin has to be antiparallel to the existing spin. After averaging over all three sites, the net effect from the addition of the fourth spin is minimized to zero.

\begin{figure}
  \centering
  \includegraphics[width=250pt]{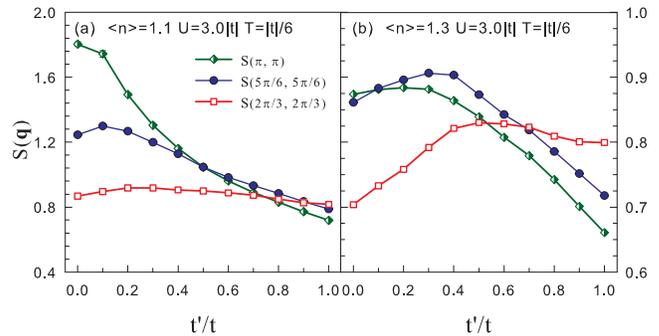}
  \caption{(color online).  Spin structure factor $S(\textbf{q})$ as a function of the frustration $t'/t$ (ranged from $t'=0$ to $t'/t=1.0$) on a $12^{2}$ lattice with $U=3.0|t|$, $T=|t|/6$ at two electron-doped fillings:(a) $\avg{n}=1.1$ and (b) $\avg{n}=1.3$. Here, three points ($\pi,\pi$) (green line), ($\frac{5\pi}{6},\frac{5\pi}{6}$) (blue line) and ($\frac{2\pi}{3},\frac{2\pi}{3}$) (red line) in the momentum space are specially concerned.}
\label{Fig:Spin12}
\end{figure}

\begin{figure}
  \centering
  \includegraphics[width=240pt]{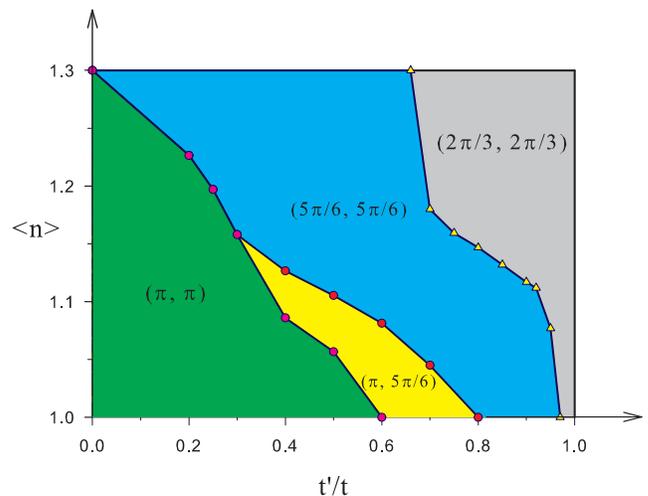}\\
  \caption{(Color online)  Magnetic phase diagram of the Hubbard model on the 2D anisotropic triangular lattice in parameter space $t'/t$ and $\avg{n}$. The results are obtained by QMC with $U=3.0|t|$ and $T=|t|/6$. The areas with different colors represent different magnetic states where ($\pi,\pi$), ($\pi,\frac{5\pi}{6}$), ($\frac{5\pi}{6},\frac{5\pi}{6}$), ($\frac{2\pi}{3},\frac{2\pi}{3}$) possess the peak and dominant position, respectively.}
\label{Fig:PD}
\end{figure}

From figures \ref{Fig:Spin}-\ref{Fig:Spinn}, it is interesting to see that there is a really rich magnetic phase diagram depends both on the frustration and
doping, and the main competition in the magnetic structure are among $S(\textbf{q})$ at ($\pi,\pi$), ($\frac{5\pi}{6},\frac{5\pi}{6}$) and ($\frac{2\pi}{3},\frac{2\pi}{3}$). To have a global picture, in Fig. \ref{Fig:Spin12}, we present the magnetic correction $S(\textbf{q})$ at these three main momentum point versus the ratio of $t'/t$ at  $\avg{n}=1.1$ (a) and  $\avg{n}=1.3$ (b).
In these two doped regions, the AFM fluctuations corresponding to $S(\pi,\pi)$ are prominent at small $t'/t$, and in contrast, $120^{\circ}$ magnetic order corresponding to $S(\frac{2\pi}{3},\frac{2\pi}{3})$ could achieve the predominance as $t'/t$ is large enough. In the between, $S(\frac{5\pi}{6},\frac{5\pi}{6})$ could take over the predominance with a moderate $t'/t$.  There are also some differences between the crossover in these two different electron filling: (i) The boundary of crossover is different, and in fact, the crossover to ($\frac{2\pi}{3},\frac{2\pi}{3}$) takes place at $t'/t=0.9$ for $\avg{n}=1.1$ (at $t'/t=0.97$ for half filling, not presented here) while the boundary between ($\frac{2\pi}{3},\frac{2\pi}{3}$) and $(\frac{5\pi}{6},\frac{5\pi}{6})$ for $\avg{n}=1.3$ is around $t'/t=0.7$, illustrating that enhancement of electron doping could make the crossover to $120^{\circ}$ ordering easier; (ii) Enhancement of the frustration promotes the spin fluctuations in $S(\frac{2\pi}{3},\frac{2\pi}{3})$ as a whole at $\avg{n}=1.3$ instead of suppressing it slightly at $\avg{n}=1.1$, which might be the reason why a sharp peak forms around the $(\frac{2\pi}{3},\frac{2\pi}{3})$ point for $t'=t$ at $\avg{n}=1.3$ (see that in Fig. \ref{Fig:Spin13}).

A full phase diagram of magnetic correlation on the $t'/t-\avg{n}$ plane is present in Fig. \ref{Fig:PD}. There is a clear crossover of magnetic order in the phase diagram, where the peak in the momentum changes along ($\pi,\pi$)$\rightarrow$($\pi,\frac{5\pi}{6}$)$\rightarrow$($\frac{5\pi}{6},\frac{5\pi}{6}$)$\rightarrow$ ($\frac{2\pi}{3},\frac{2\pi}{3}$) as $t'/t$ increases. The peak in $S(\textbf{q})$ at $(\pi,\pi)$ point appears with smaller $t'/t$ and doping, suggesting a stronger AFM fluctuation in the regions of small frustration and doping. At half filling, there is a peak at ($\frac{2\pi}{3},\frac{2\pi}{3}$) only in a very narrow region, which is close to the isotropic limit $t'=t$, indicating that $120^{\circ}$ magnetic order is stable only for $t'\sim t$ at half filling, in agreement with previous studies by other methods\cite{Gan2005,Gan2006,Powell2007,Tocchio2013}. As the system is doped away from half filling, especially at $\avg{n}=1.3$,  the $120^{\circ}$ order is dominant over a wider region of $t'/t$, comparing with that of half filling. 
 While the region where $S(\pi,\pi)$ leads, becomes more and more narrow as $\avg{n}$ increases, and the AFM correlation should be destroyed when the doping is approaching $\avg{n}=1.3$, for whichever value of $t'/t$. Thus we argue that the doping may also acts as some kind of {\it{frustration}}. We also
  note that the crossover, corresponding to ($\pi,\frac{5\pi}{6}$) and ($\frac{5\pi}{6},\frac{5\pi}{6}$), dominate in the intermediate region of moderate $t'/t$,
  suggesting the existence of one more crossover between the AFM and $120^{\circ}$ magnetic order, which possibly belongs to incommensurate $(q, q)$ long-range AFM orders\cite{Chung2001,Powell2007,Scriven2012}.  The uncertainty of the phase boundary should be in principle determined by the error bar of the magnetic correlation around the transition point, which is no more than 0.6 percent within current parameters, and the uncertainty shown in the figure is smaller than the size of symbol.

\section{Conclusion}
To summarise, we have studied the magnetic correlation for a variety of frustrations, electron fillings, interactions and temperatures within the Hubbard model on a two-dimensional anisotropic triangular lattice by using the quantum Monte Carlo method. Our non-biased numerical results provide a full phase diagram of the magnetic correlation depends on the frustration and doping. In the anisotropic triangular lattice, the system varies from non-frustrated lattice to the fully-frustrated one as $t'/t$ increases from 0 to 1.  As the  $t'/t$  increases, the peak of $S(\bf{q})$ moves from $(\pi,\pi)$, ($\pi,\frac{5\pi}{6}$), to ($\frac{5\pi}{6},\frac{5\pi}{6}$) and finally locates at ($\frac{2\pi}{3},\frac{2\pi}{3}$) at half filling. With a finite doping, the doping acts like some kinds of {\it{frustration}}, which destroys the $S(\pi,\pi)$ quickly and push the magnetic order to a wide range of $S(\frac{2\pi}{3},\frac{2\pi}{3})$ when the  $t'/t$  is large enough.  The results present here provide a rich phase diagram for the magnetic order in frustrated system.

\acknowledgements
T. Ma thanks CAEP for partial financial support. This work is supported by NSFCs (Grant. Nos. 11374034 and 11334012),
the Fundamental Research Funds for the Central Universities (Grant. No. 2014KJJCB26), and the
Fundamental Research Funds for the Central Universities in China under 2011CBA00108 (Y. Liang).

{\it Author Contributions:} Ma T. and Wang P. developed the simulation codes; Wang P., Ma X., Wang J., and Zeng Y.
performed the simulations and analyses and prepared the figures; Ma. T., Liang Y. and
Wang P. directed the investigation and wrote the paper. The manuscript reflects the
contributions of all authors.

\end{document}